\documentstyle[eqsecnum,manuscript,aps,psfig]{revtex}

\def\be{\begin{equation}}
\def\ee{\end{equation}}
\def\bea{\begin{eqnarray}}
\def\eea{\end{eqnarray}}

\begin{document}
\title{Superenergy and Supermomentum of G\"odel Universes}
\author{Mariusz P. D\c abrowski\footnote{E-mail:mpdabfz@uoo.univ.szczecin.pl} and
Janusz Garecki\footnote{E-mail:garecki@wmf.univ.szczecin.pl}}
\address{Institute of Physics, University of Szczecin, 70-451 Szczecin, Poland.}
\date{\today}
\maketitle

\begin{abstract}
We review the canonical superenergy tensor and the canonical angular supermomentum
tensors in general relativity and calculate them for space-time homogeneous
G\"odel universes to show that both of these tensors do not, in general, vanish.
We consider both an original dust-filled pressureless acausal G\"odel
model of 1949 and a scalar-field-filled causal G\"odel model of Rebou\c cas and Tiomno.
For the acausal model, the non-vanishing components of superenergy of matter are different
from those of gravitation. The angular supermomentum tensors of matter and
gravitation do not vanish either which simply reflects the fact that G\"odel
universe rotates. However, the axial (totally antisymmetric) and vectorial parts of supermomentum tensors
vanish. It is interesting that superenergetic quantities
are {\it sensitive} to causality in a way that superenergy density $_g S_{00}$
of gravitation
in the acausal model is {\it positive}, while superenergy density $_g S_{00}$ in
the causal model is {\it negative}. That means superenergetic
quantities might serve as criterion of causality in cosmology and
prove useful.

\end{abstract}


PACS number(s): 98.80.Hw

\newpage

\section{Introduction}
\label{sect1}

\setcounter{equation}{0}

In this paper we calculate the canonical superenergy tensor and the canonical
angular supermomentum tensor
for homogeneous spacetimes of the form first investigated in general relativity
by G\"odel in 1949 \cite{godel,metric}. Since we can see galaxies are now rotating there have
been suggestions that their rotation is primordial and originated from a global rotation
of an early G\"odel spacetime \cite{godel1,li}. The G\"odel's solutions
attract considerable interest because they describe rotating universes
that possess the completely unexpected property of closed timelike curves
(CTCs). However, generalized G\"odel models which do not contain CTCs
have been found in general relativity in the presence of massless
scalar fields \cite{tiomno}, and in gravity
theories derived from an action containing terms quadratic in the Ricci
curvature invariants \cite{accio}, in five-dimensional gravity theories
\cite{teixeira}, or in string-inspired gravity
theories \cite{bardab98}. We study both causal and non-causal G\"odel universes and
find their superenergetic properties.

The paper is organizing as follows. In Section II, after giving
some remarks about energy-momentum in general relativity, we
present our definition of the superenergy tensors for gravitation
and for matter. In Section III we introduce the angular supermomentum
tensors and give some hints about the application of these superenergy
and angular supermomentum tensors in general relativity. In Section IV
we present a short review of some other approaches to the problem of energy
and superenergy in general relativity.
Section V is devoted to the calculation of superenergetic quantities for
the acausal G\"odel spacetime. In Section VI the superenergetic quantities for
a generalized causal G\"odel-Rebou\c{c}as-Tiomno spacetime are given.
In Section VII we present our conclusions and in the Appendix we give some useful formulas
(for instance Bel-Robinson tensor components) to calculate superenergy and angular
supermomentum for the models under consideration.

\section{Superenergy tensors}
\label{sect2a}

\setcounter{equation}{0}

In general relativity the gravitational field which is given by Levi-Civita connection
$\Gamma^i_{kl} = \Gamma^i_{lk}$ ($i,j,a,b \ldots = 0,1,2,3$)
\footnote{Levi-Civita connection is symmetric in
holonomic coordinates.} {\it does not possess} an energy-momentum
tensor. The only objects which can be defined are {\it energy-momentum
pseudotensors} and this is a consequence of the Einstein Equivalence
Principle. Old and new investigations
(see, eg. \cite{virbhadra}) have shown that the best solution
to the energy-momentum problem in standard general relativity
(without supplementary objects like distinguished tetrad fields,
an auxilliary metric or an arbitrary vector field) seems to be the application of
the Einstein canonical energy-momentum pseudotensor $_E t_i^{~k}$
\cite{landau,moller} and the canonical double-index energy-momentum complex
\cite{landau,moller}
\be
_E K_i^{~k} = \sqrt{\vert g\vert}\bigl(T_i^{~k} + _Et_i^{~k}\bigr),
\label{complex}
\ee
where $T^{ik}$ is a symmetric energy-momentum
tensor of matter which appears on the r.h.s. of the Einstein
equations and $\vert g \vert$ is the determinant of the metric
tensor.

The Einstein equations can be rewritten in the form \cite{moller}
\begin{equation}
\sqrt{\vert g\vert}\bigl(T_i^{~k} + _E t_i^{~k}\bigr) = {_F
U_i^{~kl}}_{,l},
\label{einstein}
\end{equation}
where (in holonomic coordinates)
\begin{equation}
_F U_i^{~kl} = (-) _F U_i^{~lk} = \alpha{g_{ia}\over\sqrt{\vert
g\vert}}\bigl[(-g)\bigl(g^{ka}g^{lb} - g^{la}g^{kb}\bigr)\bigr]_{,b}
\label{freud}
\end{equation}
are Freud's superpotentials while
\begin{eqnarray}
_E t_i^{~k} &=& \alpha\Bigl\{\delta^k_i
g^{ms}\bigl(\Gamma^l_{mr}\Gamma^r_{sl} -
\Gamma^r_{ms}\Gamma^l_{rl}\bigr)\nonumber \\
&+& g^{ms}_{~~,i}\bigl[\Gamma^k_{ms} - {1\over 2}\bigl(\Gamma^k_{tp}
g^{tp} - \Gamma^l_{tl} g^{kt}\bigr)g_{ms} - {1\over
2}\bigl(\delta^k_s\Gamma^l_{ml} +
\delta^k_m\Gamma^l_{sl}\bigr)\bigr]\Bigr\},
\label{einpseudo}
\eea
is the Einstein's gravitational energy-momentum pseudotensor.
Here we have taken $(c = G = 1)$
\be
\alpha = {1 \over 16\pi} .
\ee
Since $_E t_i^{~k}$ can also be obtained from Einstein Lagrangian
of the gravitational field as a canonical object (see e.g. \cite{landau}), then
it is usually called the Einstein {\it canonical} energy-momentum
pseudotensor for the gravitational field. On the other hand, $_E K_i^{~k}$
is called the Einstein {\it canonical} energy-momentum complex for matter and
gravitation. From (\ref{einstein}) one gets the local, differential energy-momentum
conservation laws
\begin{equation}
\bigl[\sqrt{\vert g\vert}\bigl(T_i^{~k} + _E t_i^{~k}\bigr)\bigr]_{,k} =
0
\end{equation}
and the integral conservation laws (Synge's conservation laws)
\begin{equation}
\oint\limits_{\partial\Omega}{\sqrt{\vert g\vert}\bigl(T_i^{~k} + _E
t_i^{~k}\bigr)d\sigma_k} = 0,
\end{equation}
where $\partial\Omega$ is a boundary of a four-dimensional, compact
domain $\Omega$, and $d\sigma_k$ is the three-dimensional volume
element \cite{landau}.

In order to fill the gap for an energy-momentum tensor in general relativity, one
can introduce the {\it canonical
superenergy tensor} (and also other superenergy tensors) which was done
in series of papers \cite{gar96,gar99a,garjmp,gar99b}\footnote{In this paper we use the
signature $(- +++)$ despite $(+ ---)$ used in
\cite{gar96,gar99a,garjmp,gar99b}.}.
The idea of superenergy (originally introduced for the gravitational field) is quite universal
in the sense that for any physical field with an energy-momentum tensor or
pseudotensor constructed of $\Gamma^i_{kl}$ one can {\it always} build up
the corresponding superenergy tensor.

The general definition of the
superenergy tensor $S_a^{~b}(P)$ which can be applied to an arbitrary
gravitational as well as matter field is
\begin{equation}
S_{(a)}^{~~~(b)}(P) = S_a^{~b} := \displaystyle\lim_{\Omega\to
P}{\int\limits_{\Omega}\biggl[T_{(a)}^{~~~(b)}(y) -
T_{(a)}^{~~~(b)}(P)\biggr]d\Omega\over
1/2\int\limits_{\Omega}\sigma(P;y) d\Omega},
\label{averaging}
\end{equation}
where
\begin{eqnarray}
T_{(a)}^{~~~(b)}(y) &:= &
T_i^{~k}(y)e^i_{(a)}(y)e^{(b)}_k(y),\nonumber\\
T_{(a)}^{~~~(b)}(P) &:=& T_i^{~k}(P)e^i_{(a)}(P)e^{(b)}_k(P) = T_a^{~b}(P)
\end{eqnarray}
are the tetrad components of a tensor or a
pseudotensor field $T_i^{~k}(y)$ which describe an energy-momentum, $y$
is the collection of normal coordinates {\bf NC(P)} at a given point
{\bf P}, $\sigma(P,y)$ is the world-function, $e^i_{(a)}(y),~~e^{(b)}_k(y)$ denote an orthonormal tetrad
field and its dual, respectively, $e^i_{(a)}(P) =
\delta^i_a,~~e^{(a)}_k(P) = \delta^a_k$, $e^i_{(a)}(y)e^{(b)}_i(y) =
\delta^b_a$, and they are paralell propagated along
geodesics through {\bf P}. For a sufficiently small domain $\Omega$ which
surrounds {\bf P} we require
\begin{equation}
\int\limits_{\Omega}y^id\Omega = 0,~~\int\limits_{\Omega}y^iy^kd\Omega =
\delta^{ik}M,
\end{equation}
where
\begin{equation}
M = \int\limits_{\Omega}(y^0)^2d\Omega =
\int\limits_{\Omega}(y^1)^2d\Omega = \int\limits_{\Omega}(y^2)^2d\Omega
= \int\limits_{\Omega}(y^3)^2d\Omega
\end{equation}
is a common value of the moments of inertia of the domain $\Omega$
with respect to the subspaces $y^i = 0$. The procedure of an "averaging" of
the energy-momentum given in (\ref{averaging})
is an amended version of the procedure proposed by Pirani \cite{pirani}.

Let us choose $\Omega$ as a small ball defined by
\begin{equation}
(y^0)^2 + (y^1)^2 + (y^2)^2 + (y^3)^2\leq R^2,
\label{ball}
\end{equation}
which can be described in a covariant way in terms of the auxiliary
positive-definite metric $h^{ik} := 2v^iv^k + g^{ik}$, where $v^i$ are
the components of the four-velocity vector of an observer {\bf O}
at rest at {\bf P}. As for the world function we choose
\cite{synge}
\begin{equation}
\sigma(P;y){\dot =} \frac{1}{2}\bigl(-(y^0)^2 + (y^1)^2 + (y^2)^2 +
(y^3)^2\bigr)
\end{equation}
and ${\dot =}$ means it is valid only in some special coordinates.
The world function can covariantly be defined by the eikonal-like equation
\begin{equation}
g^{ik}\partial_i\sigma\partial_k\sigma = 2\sigma,
\label{eikonal}
\end{equation}
with conditions: $\sigma(P,P) = 0, ~~\partial_i\sigma(P,P) = 0$.
The equation (\ref{eikonal}) allows to rewrite (\ref{ball}) by
\begin{equation}
h^{ik}\partial_i\sigma\partial_k\sigma\leq R^2.
\end{equation}
Since at {\bf P} the tetrad and normal components are equal,
from now on we will write the components of any quantity at {\bf P}
without (tetrad) brackets, e.g., $S_a^{~b}(P)$ instead of
$S_{(a)}^{~~~(b)}(P)$ and so on.

Let us now make the following expansion for the energy-momentum tensor of matter
\cite{gar78}
\begin{eqnarray}
T_i^k(y) &=& {\hat T}_i^k + \nabla_l{\hat T}_i^k y^l +1/2 {{\hat
T}_i^k}{}_{,lm}y^ly^m +R_3\nonumber \\
&=& {\hat T}_i^k + \nabla_l {\hat T}_i^k y^l
+1/2\biggl[\nabla_{(l}\nabla_{m)} {\hat T}_i^k \nonumber \\
&-& 1/3{\hat R}^c_{~(l\vert i\vert m)} {\hat T}^k_c + 1/3 {\hat
R}^k_{~(l\vert c\vert m)} {\hat T}_i^c\biggr]y^ly^m + R_3,
\end{eqnarray}
\begin{equation}
e^i_{(a)}(y) = {\hat e}^i_{(a)} +1/6{\hat R}^i_{~lkm} {\hat
e}^k_{(a)}y^ly^m + R_3,
\label{e1}
\end{equation}
\begin{equation}
e^{(b)}_k(y) = {\hat e}^{(b)}_k - 1/6 {\hat R}^p_{~lkm}{\hat e}^{(b)}_p
y^ly^m + R_3,
\label{e2}
\end{equation}
which gives (\ref{averaging}) in the form
\begin{equation}
_m S_a^{~b}(P) = \displaystyle\lim_{\Omega\to
P}{\int\limits_{\Omega}\bigl(\nabla_l {\hat T}_a^b{}y^l
+1/2\nabla_{(l}\nabla_{m)} {\hat T}_a^b{}y^ly^m + THO\bigr)d\Omega\over
1/2\int\limits_{\Omega}\sigma(P;y)d\Omega}.
\label{avexpanded}
\end{equation}
$THO$ means the terms of higher order in the expansion of
the differences $T_{(a)}^{(b)}(y) - T_{(a)}^{(b)}(P) = T_{(a)}^{(b)}(y)
- {\hat T}_a^b$, $R_3$ is the remainder of the third order, $\nabla$ denotes covariant
differentiation, and a hat denotes the value of an object at {\bf P}.

The first and $THO$ terms in the numerator of (\ref{avexpanded}) do not
contribute to $_m S_a^b(P)$. Hence, we finally get from (\ref{avexpanded})
\begin{equation}
_m S_a^b(P) = \delta^{lm}\nabla_{(l}\nabla_{m)} T_a^b.
\label{SabTab}
\end{equation}
By introducing the four-velocity $v^l = \delta^l_0 ,~v^lv_l = -1$ of an observer {\bf
O} at rest at {\bf P} and the local metric ${\hat g}^{ab} {\dot =}
\eta^{ab}$ ($\eta^{ab}$ - Minkowski metric), one can write (\ref{SabTab}) in a covariant way as
\begin{equation}
\label{Sabmatter}
_m S_a^{~ b}(P;v^l) = \bigl(2v^lv^m + g^{lm}\bigr)
\nabla_{(l}\nabla_{m)}{} T_a^{~ b} .
\end{equation}
This is the matter superenergy tensor $_m S_a^b(P;v^l)$ and it is symmetric.
Of course, as a result of Pirani's "averaging" it does not satisfy any local conservation
laws in general relativity. However, it satisfies
trivial\footnote{They are trivial because the integral
superenergetic quantities for a closed system in special relativity
vanish.} conservation laws in special relativity.

Now let us take the gravitational field and make the expansion
\begin{eqnarray}
_E t_i^{~k}(y) &=& {\alpha\over 9}\bigl[{\hat B}^k_{~ilm} + {\hat
P}^k_{~ilm} -1/2\delta_i^k{\hat R}^{abc}_{~~~l}\bigl({\hat R}_{abcm} +
{\hat R}_{acbm}\bigr) + 2\delta_i^k{\hat R}_{(l\vert g}{}{\hat
R}^g_{~\vert m)}\nonumber \\
&-& 3 {\hat R}_{i(l\vert}{} {\hat R}^k_{~\vert m)} + 2{\hat
R}^k_{~(gi)(l\vert}{}{\hat R}^g_{~\vert m)}\bigr]y^ly^m + R_3.
\label{expgravity}
\end{eqnarray}
This expansion (\ref{expgravity}) with the help of (\ref{e1})-(\ref{e2}) gives
the {\it canonical superenergy tensor for the gravitational field}
\begin{equation}
\label{Sabgrav}
_g S_a^{~b}(P;v^l) = \bigl(2{v}^l{v}^m + {g}^{lm}\bigr) {{W}_a^{~b}}{}_{lm},
\end{equation}
where
\begin{eqnarray}
\label{Tablm}
{{W}_a^{~b}}{}_{lm} & = & {2\alpha\over 9}\bigl[{B}^b_{~alm} +
{ P}^b_{~alm} \nonumber \\
& - & {1\over 2}\delta_a^b {R}^{ijk}_{~~~m}{}\bigl({R}_{ijkl} +
{R}_{ikjl}\bigr) + 2\delta_a^b
{R}_{(l\vert g}{} {R}^g_{~\vert m)}\nonumber \\
 & - & 3 {R}_{a(l\vert}{} {R}^b_{~\vert m)}
+ 2 {R}^b_{~(ag)(l\vert}{} { R}^g_{~\vert m)}\bigr],
\end{eqnarray}
and
\begin{equation}
\label{bel}
B^b_{~alm} := 2R^{bik}_{~~~(l\vert}{} R_{aik\vert m)} - {1\over 2}\delta^b_a{}
R^{ijk}_{~~~l}{} R_{ijkm} ,
\end{equation}
is the {\it Bel--Robinson tensor}, while
\begin{equation}
\label{pel}
P^b_{~alm} := 2R^{bik}_{~~~(l\vert }{} R_{aki\vert m)} - {1\over 2}\delta_a^b
{} R^{ijk}_{~~~l}{} R_{ikjm}.
\end{equation}
In vacuum, $_g S_a^{~b}(P;v^l)$ reduces to a simpler form
\begin{equation}
_g S_a^{~b}(P;v^l) = {8\alpha\over 9}\bigl(2v^lv^m +
g^{lm}\bigr)\bigl[{R^{b(ik)}}_{(l\vert}R_{aik\vert m)}
-1/2\delta_a^b{R^{i(kp)}}_{(l\vert}R_{ikp\vert m)}\bigr],
\end{equation}
which is also symmetric and the quadratic form $_g S_{ab}(P;v^l)v^av^b$ is
positive-definite. The canonical superenergy tensor $_g S_a^{~b}(P;v^l)$ is a
tensor, contrary to the fact that the components $_E t_i^{~k}$ do not form
any geometric object. This is a consequence of the specific properties of the
normal coordinates.

Our main conjecture is that $_g S_a^{~b}(P;v^l)$
should be taken as a {\it substitute} for the energy-momentum tensor for
the gravitational field. Its advantage is that it contains the Bel-Robinson
tensor, which is a conserved quantity in vacuum.
Unfortunately, superenergy tensors do not fulfill any conservation law.
The components of $_g S_a^{~b}(P;v^l)$ and $_m S_a^{~b}(P;v^l)$ have the
dimension which can be written down as: [the dimension of the components of an
energy-momentum tensor (or pseudotensor)] $\times m^{-2}$. This might be considered
to have a straightforward physical interpretation - namely: the energy-momentum tensor
(or pseudotensor) is the {\it flux} of the appropriate superenergy tensor.
Following Bel and Robinson, we call these tensors the {\it (canonical)
superenergy tensors}.

\section{Supermomentum tensors}
\label{sect2b}

\setcounter{equation}{0}

The idea to extend the notion of superenergy onto the
angular momentum has also been proposed \cite{garjmp}.

As it is known, the notion of an angular momentum in general relativity is
much more complicated and obscure than the notion of an energy-momentum (see
e.g.\cite{winicour}). Even the matter field {\it does not possess} an
angular momentum tensor because, in general, the {\it radius vector
cannot be defined}. Moreover, one has serious difficulties with an invariant
definition of the angular momentum in an asymptotically flat spacetime
(at null or spatial infinity) \cite{winicour}, and with convergence of the resulting
global angular momentum integrals in radiative spaces. However, for a closed system,
i.e., in the case of an insular and nonradiating system \cite{moller}, one can
correctly define a global angular momentum, for example, by using Landau-Lifschitz
\cite{landau} or Bergmann-Thomson \cite{BT} angular momentum complex. In an
arbitrary asymptotically flat spacetime one can construct a reasonable formula
which gives temporal changes of the angular momentum \cite{cresswell}.

The canonical angular supermomentum tensors introduced in
\cite{garjmp} have much better geometric properties (they are tensors) than the
angular momentum complexes from which they were obtained and they lead
to global integrals which have {\it better convergence}
(at least one order better in $0(r^{-n})$) than the corresponding
global angular momenta integrals. It means that the global angular supermomenta
can also be defined in the cases where the (global) angular
momentum cannot be defined at all, i.e., when the suitable angular momentum integrals are
divergent.

The canonical angular supermomentum tensors, analogous to the case of
the canonical superenergy tensors, can be considered as substitutes for
the angular momentum tensors of matter and gravitation
which do not appear in general relativity. The constructive definition of these tensors,
in analogy to the definition of the canonical superenergy tensors, is
as follows. In normal coordinates {\bf NC(P)} we define
\begin{equation}
S^{(a)(b)(c)}(P) = S^{abc}(P) :=\displaystyle\lim_{\Omega\to
P}{\int\limits_{\Omega}\bigl[M^{(a)(b)(c)}(y) -
M^{(a)(b)(c)}(P)\bigr]d\Omega\over
1/2\int\limits_{\Omega}\sigma(P;y)d\Omega},
\label{defSabc}
\end{equation}
where
\begin{equation}
M^{(a)(b)(c)}(y) := M^{ikl}(y) e^{(a)}_i(y) e^{(b)}_k(y) e^{(c)}_l(y),
\end{equation}
\begin{eqnarray}
M^{(a)(b)(c)}(P) &:=&  M^{ikl}(P) e^{(a)}_i(P) e^{(b)}_k(P) e^{(c)}_l(P) =
M^{ikl}(P)\delta^a_i\delta^b_k\delta^c_l \nonumber \\
&=& M^{abc}(P)
\label{moment}
\end{eqnarray}
are the physical (or tetrad) components of the field $M^{ikl}(y) =
(-) M^{kil}(y)$ which describe angular momentum densities. As in (\ref{e1})-(\ref{e2}),
$e^i_{(a)}(y), ~e^{(b)}_k(y)$ denote orthonormal bases such that
$e^i_{(a)}(P) = \delta^i_a$ and its dual are parallel propagated
along geodesics through {\bf P} and $\Omega$ is a sufficiently small
ball with centre at {\bf P}. At {\bf P} the tetrad and normal
components of an object are equal. We apply this again (cf. Section \ref{sect2a})
and omit tetrad brackets for indices of
any quantity attached to the point {\bf P}; for example, we write
$S^{abc}(P)$ instead of $S^{(a)(b)(c)}(P)$ and so on.

For matter as $M^{ikl}(y)$ we take
\begin{equation}
_m M^{ikl}(y) = \sqrt{\vert g\vert}\bigl(y^i T^{kl} - y^k T^{il}\bigr),
\label{defmom}
\end{equation}
where $T^{ik}$ are the components of a symmetric
energy-momentum tensor of matter and $y^i$ denote the normal
coordinates. The formula (\ref{defmom}) gives the total angular momentum densities,
orbital and spinorial because the dynamical energy-momentum tensor
of matter $T^{ik}$ comes from the canonical energy-momentum tensor by using
the Belinfante-Rosenfeld symmetrization procedure and, therefore,
includes the spin of matter \cite{BT}.
Note that the normal coordinates $y^i$ form the components of the
local radius-vector ${\vec y}$ with respect to the origin {\bf P}. In
consequence, the components of the $_m M^{ikl}(y)$ form a tensor
density.

For the gravitational field we take the gravitational angular
momentum pseudotensor proposed by Bergmann and Thomson \cite{BT} as
\begin{equation}
_g M^{ikl}(y) = _F U^{i[kl]}(y) - _F U^{k[il]}(y) +\sqrt{\vert
g\vert}\bigl(y^i_{BT} t^{kl} - y^k_{BT} t^{il}\bigr),
\label{bergmom}
\end{equation}
where
\begin{equation}
_F U^{i[kl]} := g^{im} _FU_m^{~[kl]}
\end{equation}
are von Freud superpotentials, and
\begin{equation}
_{BT} t^{kl} := g^{ki} _E t_i^{~l} + {g^{mk}_{~~,p}\over\sqrt{\vert
g\vert}}_F U_m^{~[lp]}
\end{equation}
is the {\it Bergmann-Thomson} gravitational energy-momentum
pseudotensor. It is closely related to the canonical energy-momentum
complex and has better transformational properties than the
pseudotensor given by Landau and Lifschitz \cite{landau,gar2001}.
This is why we apply them here.

One can interpret the Bergmann-Thomson pseudotensor as the sum of the spinorial
part
\begin{equation}
S^{ikl} := _F U^{i[kl]} - _F U^{k[il]}
\end{equation}
and the orbital part
\begin{equation}
O^{ikl} := \sqrt{\vert g\vert}\bigl(y^i _{BT} t^{kl} - y^k _{BT}
t^{il}\bigr)
\end{equation}
of the gravitational angular momentum densities.

Substitution of (\ref{defmom}) and (\ref{bergmom}) (expanded up to third
order) into (\ref{defSabc}) gives the {\it canonical angular supermomentum tensors} for
matter and gravitation, respectively \cite{garjmp},
\begin{eqnarray}
\label{Sabcmatter}
_m S^{abc}(P;v^l)& = & 2\bigl[\bigl(2{v}^a {v}^p + {g}^{ap}\bigr) \nabla_p {} {T}^{bc} \nonumber \\
 & - & \bigl(2{v}^b {v}^p + {g}^{bp}\bigr)\nabla_p {} {T}^{ac}\bigr],
\end{eqnarray}
\begin{eqnarray}
\label{Sabcgrav}
_g S^{abc}(P;v^l) & = &\alpha\bigl(2{v}^p {v}^t + {g}^{pt}\bigr)\bigl[\bigl({g}^{ac} {g}^{br} - {g}^{bc} {g}^{ar}\bigr){}\nabla_{(t} {R}_{pr)} \nonumber
\\
 & + & 2{g}^{ar}\nabla_{(t}{{R}^{(b}_{~~p}}{}^{c)}_{~~r)} - 2{g}^{br} \nabla_{(t} {{R}^{(a}_{~~p}}{}^{c)}_{~~r)}\nonumber \\
& + & {2\over 3} {g}^{bc}\bigl(\nabla_r{{R}^r_{~(t}}{}^a_{~p)} -
 \nabla_{(p} {R}^a_{t)}\bigr) \nonumber \\
& - & {2\over 3}{g}^{ac}\bigl(\nabla_r {{R}^r_{~(t}}{}^b_{~p)} -
\nabla_{(p} {R}^b_{t)}\bigr)\bigr].
\end{eqnarray}
Both these tensors are antisymmetric in the first two indices $S^{abc} = - S^{bac}$.
In vacuum, the gravitational canonical angular supermomentum tensor (\ref{Sabcgrav})
simplifies to
\begin{equation}
_g S^{abc}(P;v^l) = 2\alpha\bigl(2{\hat v}^p{\hat v}^t + {\hat
g}^{pt}\bigr)\bigl[{\hat g}^{ar} \nabla_{(p}{{\hat
R}^{(b}_{~~t}}{}^{c)}_{~~r)} - {\hat g}^{br}\nabla_{(p}{{\hat
R}^{(a}_{~~t}}{}^{c)}_{~~r)}\bigr].
\end{equation}

Note that the orbital part $O^{ikl} = \sqrt{\vert g\vert}\bigl(y^i _{BT}
t^{kl} - y^k _{BT} t^{il}\bigr)$ gives no contribution to $_g S^{abc}(P;v^l)$.
Only the spinorial part $S^{ikl} = _F
U^{i[kl]} - _F U^{k[il]}$ contributes. Also, the canonical angular supermomentum tensor $_g S^{abc}(P;v^l)$
and $_m S^{abc}(P;v^l)$ of gravitation and matter do not require the introduction
of the notion of a radius vector. In special relativity the canonical angular
supermomentum tensor for matter, similarly as the canonical superenergy tensor
for matter, satisfies trivial conservation laws.

Supermomentum tensors can be decomposed into their tensor (t), vector
(v) and axial (a) (totally antisymmetric) parts as follows
\begin{equation}
\label{decomp}
S^{abc} = ^{(t)} S^{abc} + ^{(v)} S^{abc} + ^{(a)} S^{abc},
\end{equation}
where
\bea
^{(v)} S^{abc} := \frac{1}{3}\bigl(g^{bc}V^a - g^{ac} V^b\bigr),\\
^{(a)} S^{abc} = S^{[abc]} := \epsilon^{dabc}a_d,\\
V^a := S^{ab}_{~~b}, ~~a^d := - \frac{1}{6} \epsilon^{dabc}S_{abc}.
\eea
The tensorial part of $^{(t)} S^{abc}$ of a angular supermomentum tensor
can be defined as the difference $^{(t)} S^{abc} := S^{abc} -
(^{(v)} S^{abc} + ^{(a)} S^{abc})$.

The canonical superenergy tensors and the canonical
angular supermomentum tensors have successfully been calculated
for plane, plane-fronted and cylindrical gravitational waves,
Friedmann universes, Schwarzschild, and Kerr spacetimes
\cite{gar96,gar99a,garjmp,gar99b}.

The results are quite promising. For example, the superenergy
densities (which are scalars) are positive-definite for Friedman
universes and they are very useful to study the nature of the initial singularity in these
universes. Moreover, the components of the angular supermomentum tensors for
Friedman universes are equal to zero. By use of the our superenergy and
angular supermomentum tensors one can also prove that a real gravitational wave
with $R_{iklm}\not= 0$ {\it possesses and carries} positive-definite
superenergy, supermomentum and angular supermomentum, and therefore it {\it must also
have and carry} the energy-momentum and the angular momentum.
The general conclusion from our investigations in this field is that the
canonical superenergy and canonical angular supermomentum tensors give
a very useful tool for local and global analysis of the
gravitational and matter fields.

\section{Other approaches to superenergy tensors}
\label{sect2c}

\setcounter{equation}{0}

As we have already remarked we have introduced the superenergy and angular
supermomentum tensors for matter owing to universality of these tensors.
We think that these tensors are necessary for a complete description of
matter and gravity. After introducing the canonical
superenergy and angular supermomentum tensors for matter one can
define the {\it total canonical superenergy and angular supermomentum
tensors} for matter and gravitation \cite{gar96,gar99a,garjmp,gar99b} (as sums of these tensors) and
apply these total superenergy and angular supermomentum tensors to analyse a
closed system, for example. The total superenergy and angular supermomentum tensors in
an obvious way correspond to energy-momentum and angular momentum
complexes. The total canonical superenergy and total canonical angular
supermomentum tensors allow to study the {\it exchange} of superenergy
and supermomentum between gravitation and matter. Up to now, this problem
has just been studied qualitatively.

Of course, there exist other approaches to the problem of superenergy
tensors for gravitation and for matter. All of them originated from
attempts to interpret physically the Bel-Robinson
tensor\cite{bonilla,senovilla,teyssandier,senovilla2000,lazkoz}. Recent and the most thorough investigations in
this field (restricted to superenergy tensors only) have been given by Senovilla
\cite{senovilla}.

Senovilla proposes a very general and pure algebraic method for a construction
of an infinite sequence of the {\it super}$^{(k)}$--{\it energy tensors}
$(k = 1,2,3,...)$ for any linear physical field $\Psi$. This
method is independent of the field equations and formalism of the
canonical energy-momentum. It is a formal generalization (onto any
physical field $\Psi$ which satisfies linear field equations and onto
its covariant derivatives of an arbitrary order) of the algebraic method
of the construction of the symmetric energy-momentum tensor for
electromagnetic field and of the Bel-Robinson and Bel
tensors\footnote{In vacuum these two tensors coincide.}. Of course, a
symmetric energy-momentum of the field $\Psi$, the symmetric
energy-momentum tensor for the electromagnetic field, the Bel-Robinson
and Bel tensors are included in the infinite sequences of the super$^{(k)}$
--energy tensors. For example, the Bel-Robinson tensor is the
super$^{(1)}$ --energy-tensor for the Weyl curvature tensor field and the
symmetric energy-momentum tensor for electromagnetic field is
super$^{(2)}$ --energy tensor for this field.

A general method of construction of the super$^{(k)}$--energy
tensors $(k = 1,2,3,...)$ for a linear field (but only in special
relativity), similar to the method given in \cite{senovilla}, was also proposed
by Teyssandier \cite{teyssandier}.

It is difficult to find a link between our superenergy tensors and
the tensors given in \cite{senovilla} (or in \cite{teyssandier}). One can only say that in general
our superenergy tensors correspond, in some sense, to those superenergy
tensors given in \cite{senovilla} (or in \cite{teyssandier}) which are
constructed from derivatives of the third order of a physical field $\Psi$,i.e., {\it our
superenergy tensors correspond to the super}$^{(4)}${\it  - energy tensors} of
Senovilla (or Teyssandier).

Our approach seems to be much simpler than the approach developed in \cite{senovilla} (or in \cite{teyssandier}) and it has profound physical meaning: our
superenergy (and supermomentum) tensors {\it are  simply the tensors of an
averaged relative energy-momentum} (and relative angular momentum). In
gravitational case our superenergy and angular supermomentum tensors {\it
extract} covariant information about gravitational field which is hidden in
gravitational pseudotensors.

In special relativity our superenergy and supermomentum tensors {\it do not introduce} any
new intrinsic conservation laws for a closed system apart from those which are satisfied
by the  symmetric energy-momentum tensor. On the other hand, the approach
developed in \cite{senovilla} (or in \cite{teyssandier}) is far from the standard formalism
of the energy-momentum, e.g., in this approach one can try to give a physical
meaning to infinite sequence of derivatives of a linear physical field
$\Psi$ and in special relativity this approach  leads to an infinitely many
conservation laws for such a field and its derivatives. From these
conservation laws, in fact, only the 10 conservation laws, satisfied by the
symmetric energy-momentum tensor of $\Psi$, can be physically valid.

Our superenergy and angular supermomentum tensors satisfy only 10 local
conservation laws in special relativity as a consequence of the 10 conservation
laws which are satisfied by the symmetric energy-momentum tensor of
matter. However, these local conservation laws {\it are trivial} in the sense
that {\it they do not lead} to any new integral conservation laws for a closed
system because the integral superenergetic quantities calculated from
our superenergy and angular supermomentum tensors are all equal
to zero for a closed system. So, we have only 10 intrinsic
conservation laws which are satisfied by the symmetric energy-momentum
tensor, i.e., exactly the number of conservation laws required in
special relativity.

It seems to us that one needs also some deeper physical interpretation to
super$^{(k)}$--energy tensors $(k = 1,2,3,...)$ (except those
which are simply symmetric energy-momentum tensors of the appropriate
fields). This is because all the physical content of any physical field
$\Psi$ in general relativity is contained in Einstein equations with the
symmetric energy-momentum of this field as sources and in the field
equations which are satisfied by the field $\Psi$.

Resuming, we think that our definition of the superenergy (and angular
supermomentum) tensors in general relativity is very useful from practical point
of view and has a good physical motivation. In consequence, we will
confine in the following to our approach.

\section{Superenergetic quantities for acausal G\"odel spacetime}
\label{sect3}
\setcounter{equation}{0}

The G\"odel metric describes a space-time homogeneous, but anisotropic universe \cite{godel}.
There exists a five-dimensional group of isometries which acts in G\"odel universe and it is transitive.
Its line element in cylindrical coordinates $(x^0,x^1,x^2,x^3) \equiv (t,r,\psi,z)$ in an
orthonormal frame is given by \cite{tiomno}
\begin{equation}
\label{godelmet}
ds^2=-(e^0)^2 + (e^1)^2 + (e^2)^2 + (e^3)^2,
\end{equation}
where
\begin{eqnarray}
\label{eeee}
e^0 & = & dt + H(r) d\psi \nonumber \\
e^1 & = & dr \nonumber \\
e^2 & = & D(r) d\psi \nonumber \\
e^3 & = & dz
\end{eqnarray}
and the radial functions have the form
\begin{equation}
\label{HaDe}
H(r) = \sqrt{2} D(r) = e^{\sqrt{2} \Omega r}
\end{equation}
with $\Omega $ constant.
In a G\"odel universe, the four-velocity of matter is $u^l =\delta
_0^l $ and the rotation vector is $V^l =\Omega \delta^l_3$, while
the vorticity scalar is given by $\omega =\Omega /\sqrt{2}$. The
G\"odel metric (\ref{godelmet}) is a solution to the Einstein's
field equations for matter containing dust with constant energy
density $\varrho$ and the (negative) cosmological constant. The
energy-momentum tensor is \cite{visser}
\be
T_{ab} = T^{(d)}_{ab} + T^{\Lambda}_{ab} ,
\ee
where
\bea
\label{enmom}
T^{(d)}_{ab} & = & \varrho v_a v_b ,\\
\label{TLambda}
T^{(\Lambda)}_{ab} & = & - {\Lambda\over 8\pi} \eta_{ab} ,
\eea
or, explicitely,
\bea
\label{Tabexplicit}
T^{(d)}_{00} & = & \varrho,\hspace{0.5cm} T^{(d)}_{11} = T^{(d)}_{22} = T^{(d)}_{33}
= 0 \\
T^{(\Lambda)}_{00} & = & -T^{(\Lambda)}_{11} = - T^{(\Lambda)}_{22} = - T^{(\Lambda)}_{33} =
{\Lambda\over 8\pi} ,
\eea
and the following relation must be fulfilled
\begin{equation}
\label{rho}
4\pi \varrho = \Omega^2 = - \Lambda = {\rm const.}
\end{equation}

The unexpected property of the metric (\ref{godelmet}) with $H$
and $D$ given by (\ref{HaDe}) is that it permits time travel,
i.e., there exists a closed timelike curve (CTC) through every
point of spacetime \cite{metric}. In other words, this spacetime
contains a closed chronological curve and so a
chronology-violating time machine \cite{visser}. A time machine is
an object or a system which permits travel into the past - this
leads to a paradox, since one is then also able to influence one's
own future (which is also one's past). There exist chronology-violating
time machines and causality-violating time machines (those which allows either timelike
or null closed curves). Chronology violation implies causality violation and
this is why we speak in this section about an acausal G\"odel model. Another problem is that
CTCs make it impossible to foliate G\"odel spacetime into
spacelike hypersurfaces, so that Cauchy problem is ill-posed since
one cannot say what are the "initial data" which evolve (in fact,
these data do not exist at all). This also means there exists no
global cosmic time coordinate $t$, despite the fact that the G\"odel
spacetime is geodesically complete.

Since even the simple Minkowski spacetime can be made to contain
CTCs by simply identifying points with $t = 0$ and $t = T$ for $t \in
[0,T]$ which nobody believes it is acceptable, then most of the
physicists consider G\"odel model as unphysical \cite{visser}. This was best
expressed in terms of the Hawking's chronology protection
conjecture \cite{hawking} which says that time travel is
completely forbidden in the universe. As we shall see in Section
\ref{sect4} one is able to find a generalized G\"odel model
which is chronology-protected.

Because in acausal G\"odel universe the spatial hypersurfaces $t = const$ do not
exist we confine only to the local analysis of the
superenergy and angular supermomentum in this universe.

The only nonvanishing components of the Ricci rotation coefficients $\gamma^k_{ab}$
for the metric (\ref{godelmet}) are given in the Appendix while the Riemann tensor components
in an orthonormal frame permitted by the spacetime homogeneity of the G\"odel universe
are \cite{tiomno,accio}
\begin{equation}
R_{0101}=R_{0202}=\frac 14\left( \frac{H^{\prime }}D\right) ^2=\Omega ^2,%
\hspace{0.5cm}R_{1212}=\frac 34\left( \frac{H^{\prime }}D\right)^2 -\frac{%
D^{\prime \prime }}D=\Omega^2,
\end{equation}
and the prime means the derivative with respect to $r$. The
nonzero components of the Ricci tensor and the Ricci scalar read as
\begin{equation}
R_{00} = 2\Omega^2, \hspace{0.5cm} R_{11} = R_{22} = 0,
\hspace{0.5cm} R = - 2\Omega^2   .
\end{equation}
In this paper we take the natural orthonormal frame (\ref{godelmet}) for G\"odel spacetime
as the basic tetrads of the {\bf NC(P)} ({\bf P} -- variable). In consequence,
we have $v^l = u^l = \delta^l_0$ and $g^{ab}({\bf P}) = \eta^{ab}$.

First let us calculate the superenergy of matter. From
(\ref{Sabmatter}) we have for the metric (\ref{godelmet})
\be
\label{TTTT}
_m S_{ab} = T_{ab;(00)} + T_{ab;(11)} + T_{ab;(22)} + T_{ab;(33)},
\ee
where we have replaced $\nabla(\ldots)$ by $(\ldots)_{;}$.
Since
\bea
T^{(\Lambda)}_{ma;bc}  =  (- \Lambda) \eta_{ma;bc} =
 (- \Lambda) \left\{ \left[ \eta_{ma,b} - \gamma^{k}_{mb}
\eta_{ka} - \gamma^{k}_{ab} \eta_{mk} \right]_{,c}
 -  \gamma^{k}_{mc} \left( \eta_{ka,b} - \gamma^{l}_{kb} \eta_{la} -
\gamma^{l}_{ab} \eta_{kl} \right) \right. \nonumber \\ \left.
 -  \gamma^{k}_{ac} \left( \eta_{mk,b} - \gamma^{l}_{mb} \eta_{lk} -
\gamma^{l}_{kb} \eta_{ml} \right)
 -  \gamma^{k}_{bc} \left( \eta_{ma,k} - \gamma^{l}_{mk} \eta_{la} -
\gamma^{l}_{ak} \eta_{ml} \right) \right\},
\eea
then applying the appropriate coefficients $\gamma^{a}_{bc}$ for metric
(\ref{godelmet}) (see Appendix) one can easily show that the
cosmological constant {\it does not contribute} to superenergy at
all, i.e.,
\be
_m S^{(\Lambda)}_{ab} = 0  .
\ee
However, for the dust we have
\bea
T^{(d)}_{ma;bc}  =  \left[ T^{(d)}_{ma,b} - \gamma^{k}_{mb}
T^{(d)}_{ka} - \gamma^{k}_{ab} T^{(d)}_{mk} \right]_{,c}
 -  \gamma^{k}_{mc} \left( T^{(d)}_{ka,b} - \gamma^{l}_{kb} T^{(d)}_{la} -
\gamma^{l}_{ab} T^{(d)}_{kl} \right) \nonumber \\
 -  \gamma^{k}_{ac} \left( T^{(d)}_{mk,b} - \gamma^{l}_{mb} T^{(d)}_{lk} -
\gamma^{l}_{kb} T^{(d)}_{ml} \right)
 -  \gamma^{k}_{bc} \left( T^{(d)}_{ma,k} - \gamma^{l}_{mk} T^{(d)}_{la} -
\gamma^{l}_{ak} T^{(d)}_{ml} \right),
\eea
and we calculate that it contributes to the superenergy of matter,
and the canonical superenergy tensor (\ref{Sabmatter}) for matter fields
read as
\begin{equation}
\label{godmS00}
_{m}S_0^0 = - \frac{\Omega^4}{\pi}, \hspace{0.5cm}_{m}S_1^1 =_{m}S_2^2 =
 \frac{\Omega^4}{2\pi}.
\end{equation}
It is obvious from (\ref{godmS00}) that the superenergy tensor of
matter for G\"odel models is traceless.

For the gravitational field, the non-vanishing components of the canonical
superenergy tensor (\ref{Sabgrav}) are
\begin{equation}
\label{godgS00}
_gS_0^0 = - \frac{\Omega^4}{36\pi} , \hspace{0.5cm} _gS_1^1 = _gS_2^2 = - \frac{\Omega^4}{9\pi}
, \hspace{0.5cm} _gS_3^3 = - \frac{7\Omega^4}{36\pi} .
\end{equation}
From (\ref{godgS00}) we notice that superenergy density $S_{00}$
is positive.

On the other hand, the components of the canonical angular supermomentum tensors for
matter fields (\ref{Sabcmatter}) are
\begin{equation}
\label{mSabc}
_{m}S^{012} = \frac{\Omega^3}{2\pi}, \hspace{0.5cm}
_{m}S^{201} = \frac{\Omega^3}{2\pi}, \hspace{0.5cm}
_{m}S^{120} = - \frac{\Omega^3}{\pi},
\end{equation}
while for the gravitational field (\ref{Sabcgrav}) one has
\begin{equation}
\label{gSabc}
_{g}S^{012} = \frac{4\Omega^3}{24\pi}, \hspace{0.5cm}
_{g}S^{201} = \frac{13\Omega^3}{24\pi}, \hspace{0.5cm}
_{g}S^{120} = - \frac{17\Omega^3}{24\pi}.
\end{equation}
It is easy to notice that axial (totally antisymmetric) parts of both $_{m}S^{abc}$
and $_{g}S^{abc}$ given by equations (\ref{mSabc}) and (\ref{gSabc}) for G\"odel
model (\ref{godelmet}) vanish, i.e.,
the antisymmetric part
\begin{equation}
^{(a)} S^{abc} = S^{[abc]} = \epsilon^{3abc}A_{3} = 0  ,
\end{equation}
where
\be
A_3 = - \frac{1}{6} \epsilon_{3abc} S^{abc}  .
\ee
This is also the case for the vector part of supermomentum tensor $^{(V)} S^{abc} =
0$, so that only pure tensorial part remains non-vanishing (see Eq. (\ref{decomp})).

\section{Superenergetic quantities for causal G\"odel spacetime}
\label{sect4}
\setcounter{equation}{0}

A completely causal G\"odel universe with matter source being the
scalar field without potential and (negative) cosmological constant has been found by Rebou\c cas and
Tiomno \cite{tiomno}. Its metric can be written in the form
(\ref{godelmet}) provided we replace $H(r)$ and $D(r)$ into the
following generalized functions of the radial coordinate
\begin{eqnarray}
\label{HaDec}
H(r) &=&\frac{4\Omega }{m^2}\sinh ^2{\left( \frac{mr}2\right) }, \\
D(r) &=&\frac 1{m}\sinh {(mr)}.
\end{eqnarray}
This allows to write the metric (\ref{godelmet}) in the form
\be
\label{godelmetc}
ds^2=-dt^2-2H(r)dtd\psi +G(r)d\psi ^2+dr^2+dz^2,
\ee
where
\be
G(r) =\frac 4{m^2}\sinh ^2{\left( \frac{mr}2\right) }\left[ 1+\left( 1-
\frac{4\Omega ^2}{m^2}\right) \sinh ^2{\left( \frac{mr}2\right) }\right] ,
\ee
with $m$ and $\Omega $ constants. This model is causal (there are no CTCs) because
\begin{equation}
\label{causcond}
G(r)=D^2(r)-H^2(r)>0
\end{equation}
for
\be
\label{RT}
4\Omega^2 = m^2,
\ee
and the term in front of $d\psi^2$ in the metric (\ref{godelmetc})
remains positive. The conditions (\ref{causcond}) and (\ref{RT}) remove CTCs to a point
which is formally at $r \to \infty$. On the other hand, the function $H(r)$ can always
be made zero in a more general
class of models studied in Refs. \cite{carrion,Reb98} for which there are no CTCs for
any value of the radial coordinate $r > 0$.  The G\"odel model (\ref{godelmet}) is obtained
by taking
\be
\label{G}
2\Omega^2 = m^2
\ee
in (\ref{HaDe}) and it does not fulfil the condition (\ref{causcond}).

The only nonvanishing components of the Riemann tensor for the metric (\ref{godelmetc})
in an orthonormal frame are \cite{tiomno,accio}
\begin{equation}
R_{0101}=R_{0202}=\frac 14\left( \frac{H^{\prime }}D\right) ^2=\Omega ^2,
\hspace{0.5cm}R_{1212}=\frac 34\left( \frac{H^{\prime }}D\right)^2 -\frac{
D^{\prime \prime }}D=- \Omega^2.
\end{equation}
The nonzero components of the Ricci tensor and the Ricci scalar read as
\begin{equation}
R_{00} = 2\Omega^2, \hspace{0.5cm} R_{11} = R_{22} = -2 \Omega^2,
\hspace{0.5cm} R = - 6\Omega^2   .
\end{equation}
The scalar field depends
only on the coordinate along the axis of rotation, $z$, i.e.,
\begin{equation}
\phi =\phi (z)=e z+\phi _0,
\end{equation}
where $e$ and $\phi _0$ are constants. The energy-momentum tensor of the scalar field reads
as
\be
\label{Tphi}
T^{(\phi)}_{ab} = \phi_{;a}\phi_{;b} - \frac{1}{2} \eta_{ab} \phi_{;m} \phi^{;m}
,
\ee
so because $\phi_{;a} = e \delta_{a}^{3}$, then one has
\be
T_{00} = -T_{11} = - T_{22} = T_{33} = \frac{1}{2} e^2  .
\ee
For the cosmological term we have the
energy-momentum tensor given again by (\ref{TLambda}).
The following relation between parameters $\Lambda, \Omega$ and $e$ has to be fulfilled
\be
\Lambda = -2 \Omega^2 = - 8\pi e^2 .
\ee

After simple, but tedious calculations (see Appendix) we have found that all
components of the canonical superenergy tensor (\ref{Sabmatter}) for matter fields
(\ref{Tphi})-(\ref{TLambda}) vanish (we have already shown in the Section \ref{sect3} that
this is the case for the cosmological constant), i.e.
\begin{equation}
\label{godmS00caus}
_{m}S_a^b = 0
\end{equation}

For the gravitational field, the non-vanishing components of the canonical
superenergy tensor (\ref{Sabgrav}) are
\begin{equation}
\label{gS00caus}
_gS_0^0 = \frac{\Omega^4}{4\pi} , \hspace{0.5cm} _gS_1^1 = _gS_2^2 = - \frac{\Omega^4}{36\pi}
, \hspace{0.5cm} _gS_3^3 = - \frac{\Omega^4}{12\pi} .
\end{equation}
From (\ref{gS00caus}) it follows that the price to pay for
causality is the negative superenergy density $_gS_{00} < 0$.

On the other hand, the components of the canonical angular supermomentum tensors for
matter fields (\ref{Sabcmatter}) vanish
\begin{equation}
\label{mSabc1}
_{m}S^{abc} = 0 ,
\end{equation}
while for the gravitational field (\ref{Sabcgrav}) one has
\begin{equation}
\label{gSabc1}
_{g}S^{012} = 0, \hspace{0.5cm}
_{g}S^{201} = \frac{\sqrt{2}\Omega^3}{24\pi}, \hspace{0.5cm}
_{g}S^{120} = - \frac{\sqrt{2}\Omega^3}{24\pi}.
\end{equation}
It is easy to notice that $_{g}S^{abc}$
given by (\ref{gSabc1}) for G\"odel
model (\ref{godelmetc}) is axial-free and vector-free (cf. Eq. (\ref{decomp}), i.e.,
\bea
^{(a)}S^{[abc]} &=& \epsilon^{3abc}A_{3} = 0  ,\\
^{(V)} S^{abc} &=& 0  .
\eea

\section{Conclusion}

We have calculated the canonical superenergy tensor and the canonical angular supermomentum
tensor for homogeneous rotating G\"odel universes. We considered the original pressureless
dust plus dosmological constant model found by G\"odel in 1949 which admits CTCs (acausal model) and the
scalar-field plus cosmological constant model found by Rebou\c cas and Tiomno in 1983
which is free from CTCs (causal model). Due to the peculiarity of G\"odel
spacetimes we expected the appearance of some interesting
properties of the calculated superenergetic quantities. On the other hand, because
of rotation we expected the appearance of some non-vanishing
components of the angular supermomentum tensor. For the acausal model the non-vanishing
components of superenergy of matter are different
from those of gravitation. The matter superenergy tensor is traceless.
The angular supermomentum tensors of matter and
gravitation do not vanish either which simply reflects the fact that G\"odel
universe rotates. It emerges that the supermomentum tensors have vanishing totally
antisymmetric (axial) and vectorial parts. For the causal model superenergy
and supermomentum of matter vanish. However, superenergy and
supermomentum of gravitation do not vanish. On the other hand, superenergy density
for the causal model is negative and its supermomentum is axial-free and vector-free as for
the acausal model.

It is interesting that superenergetic quantities
are {\it sensitive} to causality in a way that superenergy density $_g S_{00}$
of gravitation
in the acausal model is {\it positive}, while superenergy density $_g S_{00}$ in
the causal model is {\it negative}. That means superenergetic
quantities might serve as criterion of causality in cosmology and
prove useful.

Another point is that although both G\"odel-type models
have as a source of gravity the cosmological constant it does not
contribute to their superenergy and supermomentum at all.

For our requirements it is important that in the acausal G\"odel
universe does not exist a global cosmic time coordinate $t$. This
is the reason why we confine only to local analysis of the
superenergy and the angular supermomentum in this universe.
However, it is different in a causal G\"odel-Rebou\c{c}as-Tiomno universe
where the global hypersurfaces $t = const$. exist, so it would be
possible to calculate the energetic quantities, i.e., the energy, the momentum and the
angular momentum for these universes and compare the results with superenergetic
quantities. Such analysis has already been given for Friedman
universes \cite{gar99a,garjmp}.

Finally, we believe that superenergetic quantities give quite a lot of
physical information about the specific properties of G\"odel
spacetime which make these quantities useful.

\section{Acknowledgments}

The authors would like to thank anonymous Referees for their comments and suggestions.

\appendix

\section{Useful formulas to calculate superenergy and
supermomentum}

\setcounter{equation}{0}

The Ricci rotation coefficients for a generalized G\"odel model
(\ref{godelmetc}) are
\bea
\gamma^{0}_{12} = \gamma^{1}_{20} = \gamma^{1}_{02} =
\frac{m}{\sqrt{2}}, \nonumber \\
\gamma^{0}_{21} = \gamma^{2}_{10} = \gamma^{2}_{01} =
- \frac{m}{\sqrt{2}}, \nonumber \\
\gamma^{1}_{22} = - \gamma^{2}_{12} = - m .
\eea
According to (\ref{RT}) and (\ref{G}) one has to put $m = \sqrt{2} \Omega$
for G\"odel model and $m = 2\Omega$ for Rebou\c{c}as and Tiomno model.

The nonzero components of the Bel-Robinson tensor (\ref{bel}) are
\bea
B^0\hspace{0pt}_{00}\hspace{0 pt}^0 & = & - B^3\hspace{0pt}_{30}\hspace{0 pt}^0  = 2
\Omega^4 ,
\nonumber \\
B^0\hspace{0pt}_{01}\hspace{0 pt}^1 & = & B^0\hspace{0pt}_{02}\hspace{0pt}^2
= B^0\hspace{0pt}_{03}\hspace{0 pt}^3 = - B^1\hspace{0pt}_{12}\hspace{0 pt}^2 =
- B^2\hspace{0pt}_{21}\hspace{0 pt}^1 = -8 \Omega^4 + 6 \Omega^2 m^2 - m^4
,
\nonumber \\
B^2\hspace{0pt}_{22}\hspace{0 pt}^2 & = & B^1\hspace{0pt}_{11}\hspace{0pt}^1
= - B^3\hspace{0pt}_{31}\hspace{0 pt}^1 = - B^3\hspace{0pt}_{32}\hspace{0 pt}^2
= 10 \Omega^4 - 6 \Omega^2 m^2 + m^4 .
\eea
The non-vanishing components of the tensor (\ref{pel}) are
\bea
P^0\hspace{0pt}_{00}\hspace{0 pt}^0 & = & 3 \Omega^4, \hspace{1.cm}
P^1\hspace{0pt}_{10}\hspace{0 pt}^0  = P^2\hspace{0pt}_{20}\hspace{0 pt}^0  =
- P^3\hspace{0pt}_{30}\hspace{0 pt}^0 = - P^3\hspace{0pt}_{31}\hspace{0 pt}^1 =
- P^3\hspace{0pt}_{32}\hspace{0 pt}^2 =  \Omega^4 ,
\nonumber \\
P^0\hspace{0pt}_{01}\hspace{0 pt}^1 & = & P^0\hspace{0pt}_{02}\hspace{0 pt}^2
= P^0\hspace{0pt}_{03}\hspace{0 pt}^3 = - \frac{1}{2} \left( 6
\Omega^4 - 6 \Omega^2 m^2 + m^4 \right) , \nonumber \\
P^1\hspace{0pt}_{11}\hspace{0 pt}^1 & = & P^2\hspace{0pt}_{22}\hspace{0 pt}^2
= \frac{3}{2} \left( 10 \Omega^4 - 6 \Omega^2 m^2 + m^4 \right) ,
\nonumber \\
P^1\hspace{0pt}_{12}\hspace{0 pt}^2 & = &  P^2\hspace{0pt}_{21}\hspace{0 pt}^1
= \frac{1}{2} \left( 26 \Omega^4 - 18 \Omega^2 m^2 + m^4 \right) .
\eea
We also define (see Eq. (\ref{Tablm}))
\be
Riem2^l_l :=  -  {1\over 2} {R}^{ijk}_{~~~m}{}\bigl({R}_{ijkl} +
{R}_{ikjl}\bigr) ,
\ee
which has non-vanishing components
\be
Riem2^0_0 = - 3 \Omega^4 , \hspace{1.cm} Riem2^1_1 = Riem2^2_2 = -
\frac{3}{2} \left( 10 \Omega^4 - 6 \Omega^2 m^2 + m^4 \right) .
\ee
Another object which we define is
\be
_a^a Ric_{ll} := 2\delta_a^b
{R}_{(l\vert g}{} {R}^g_{~\vert l)}
  -  3 {R}_{a(l\vert}{} {R}^b_{~\vert l)}
+ 2 {R}^b_{~(ag)(l\vert}{} { R}^g_{~\vert l)} ,
\ee
and its nonzero components are
\bea
_0^0 Ric^0_0 & = & - 4 \Omega^4 , \hspace{1.cm} _1^1 Ric^0_0 = _2^2 Ric^0_0
= 10 \Omega^4 , \hspace{1.cm} _3^3 Ric^0_0 = 8 \Omega^4 ,\nonumber \\
_0^0 Ric^1_1 & = & _0^0 Ric^2_2 = 6 \Omega^4 - 7 \Omega^2 m^2 +
2m^4 , \hspace{1.cm} _1^1 Ric^2_2 = _2^2 Ric^1_1 = 14 \Omega^4 - 13
\Omega^2 m^2 + 3m^4 , \nonumber \\
_1^1 Ric^1_1 & = & _2^2 Ric^2_2 = -4\Omega^4 + 4 \Omega^2 m^2 -
m^4 , \hspace{1.cm} _3^3 Ric^1_1 = _3^3 Ric^2_2 = \left( 2 \Omega^2
- m^2 \right)^2  .
\eea
Finally, the non-vanishing components of the superenergy tensor
(\ref{Sabgrav}) are
\bea
_g S^1_1 & = & _g S^2_2 = \frac{2\alpha}{9} \left[ 18 \Omega^4 -
21 \Omega^2 m^2 + 4 m^4 \right] , \nonumber \\
g S^3_3 & = & \frac{2\alpha}{9} \left[ - 30 \Omega^4 +
10 \Omega^2 m^2 - m^4 \right] ,\nonumber \\
g S^0_0 & = & \frac{2\alpha}{9} \left[ -38 \Omega^4 +
22 \Omega^2 m^2 - 2 m^4 \right] .
\eea
In order to calculate supermomentum of gravity we need covariant
derivatives of the Riemann tensor
\be
R^b\hspace{0 pt}_p\hspace{0 pt}^c\hspace{0 pt}_{r;t} = R^b\hspace{0 pt}_p\hspace{0 pt}^c\hspace{0 pt}_{r,t} +
\gamma^b_{kt} R^k\hspace{0 pt}_p\hspace{0 pt}^c\hspace{0 pt}_r - \gamma^k_{pt} R^b\hspace{0 pt}_k\hspace{0 pt}^c\hspace{0 pt}_r
+ \gamma^c_{kt} R^b\hspace{0 pt}_p\hspace{0 pt}^k\hspace{0 pt}_r - \gamma^k_{rt} R^b\hspace{0 pt}_p\hspace{0 pt}^c\hspace{0 pt}_k
,
\ee
which have nonzero components
\bea
R^2\hspace{0 pt}_0\hspace{0 pt}^1\hspace{0 pt}_{0;0} & = &
R^1\hspace{0 pt}_0\hspace{0 pt}^2\hspace{0 pt}_{0;0} =
\frac{m}{\sqrt{2}} \Omega^2 ,
\hspace{1.cm} R^2\hspace{0 pt}_1\hspace{0 pt}^1\hspace{0 pt}_{0;1} =
R^1\hspace{0 pt}_0\hspace{0 pt}^2\hspace{0 pt}_{1;1} = - \frac{m}{\sqrt{2}} \left(
-2 \Omega^2 + m^2 \right) ,\nonumber \\
R^1\hspace{0 pt}_2\hspace{0 pt}^2\hspace{0 pt}_{0;2} & = &
R^2\hspace{0 pt}_0\hspace{0 pt}^1\hspace{0 pt}_{2;2} =
R^0\hspace{0 pt}_2\hspace{0 pt}^1\hspace{0 pt}_{2;2} =
R^1\hspace{0 pt}_2\hspace{0 pt}^0\hspace{0 pt}_{2;2} =
R^0\hspace{0 pt}_1\hspace{0 pt}^1\hspace{0 pt}_{2;1} =
R^1\hspace{0 pt}_2\hspace{0 pt}^0\hspace{0 pt}_{1;1} =
\frac{m}{\sqrt{2}} \left( 4 \Omega^2 - m^2 \right) ,
\nonumber \\
R^0\hspace{0 pt}_2\hspace{0 pt}^2\hspace{0 pt}_{1;2} & = &
R^2\hspace{0 pt}_1\hspace{0 pt}^0\hspace{0 pt}_{2;2} =
R^0\hspace{0 pt}_1\hspace{0 pt}^2\hspace{0 pt}_{1;1} =
R^2\hspace{0 pt}_1\hspace{0 pt}^0\hspace{0 pt}_{1;1} =
- \frac{m}{\sqrt{2}} \left( 4 \Omega^2 - m^2 \right) ,
\eea
and finally for (\ref{Sabcgrav}) we have
\bea
\frac{1}{\alpha}_g S^{012} & = &  \frac{2\sqrt{2}m}{3} \left( 4 \Omega^2 - m^2
\right) ,\nonumber \\
\frac{1}{\alpha}_g S^{021} & = &  \frac{\sqrt{2}m}{3} \left( -25 \Omega^2 + 6m^2
\right) , \nonumber \\
\frac{1}{\alpha}_g S^{120} & = &  \frac{\sqrt{2}m}{3} \left( -33 \Omega^2 + 8m^2
\right) .
\eea

\end{document}